\documentclass[aps,prc,twocolumn,showpacs,superscriptaddress,longbibliography,floatfix,10pt]{revtex4-1}

\usepackage{bm}
\usepackage{dcolumn}
\usepackage{amssymb}
\usepackage{amsmath}
\usepackage{graphicx}
\usepackage{dcolumn}
\usepackage{bm}
\usepackage{xcolor}
\usepackage{fix-cm}
\usepackage{mathptmx} 
\usepackage[T1]{fontenc}
\usepackage[colorlinks,citecolor=blue,linkcolor=blue,anchorcolor=blue,filecolor=blue,urlcolor=blue]{hyperref}
\usepackage{multirow}
\usepackage{CJK}
\makeatletter
\def\NAT@def@citea{\def\@citea{\NAT@separator}}
\makeatother

\begin{document}

\title{Influence of the tensor interaction on heavy-ion fusion cross sections}
\begin{CJK*}{UTF8}{gbsn}
\author{K. Godbey} \email{kyle.s.godbey@vanderbilt.edu}
\affiliation{Department of Physics and Astronomy, Vanderbilt University, Nashville, TN 37235, USA}
\author{Lu Guo (郭璐)} \email{luguo@ucas.ac.cn}
\affiliation{School of Nuclear Science and Technology, University of Chinese Academy of Sciences, Beijing 100049, China}
\affiliation{Institute of Theoretical Physics, Chinese Academy of Sciences, Beijing 100190, China}
\author{A. S. Umar} \email{umar@compsci.cas.vanderbilt.edu}
\affiliation{Department of Physics and Astronomy, Vanderbilt University, Nashville, TN 37235, USA}

\date{\today}

\begin{abstract}
\edef\oldrightskip{\the\rightskip}
\begin{description}
        \rightskip\oldrightskip\relax
        \setlength{\parskip}{0pt}
\item[Background]

While the tensor interaction has been shown to significantly affect the nuclear structure of exotic nuclei, its influence on nuclear reactions has only recently been investigated.
The primary reason for this neglect is the fact that most studies of nuclear dynamics do not include the tensor force at all in their models.
Indeed, only a few Skyrme parametrizations consider the tensor interaction in parameter determination.
With modern research facilities extending our ability to probe exotic nuclei, a correct description of nuclear dynamics and heavy-ion fusion is vital to both supporting and leading experimental efforts.

\item[Purpose]
To investigate the effect of the tensor interaction on fusion cross sections for a variety of nuclear reactions spanning light and heavy nuclei.

\item[Method]
Fusion cross sections are calculated using ion-ion potentials generated by the fully microscopic density-constrained time-dependent Hartree-Fock (DC-TDHF) method with the complete Skyrme tensor interaction.

\item[Results]
For light nuclei, the tensor force only slightly changes the sub-barrier fusion cross sections at very low energies.
Heavier nuclei, however, begin to exhibit a substantial hindrance effect in the sub-barrier region.
This effect is strongest in spin-unsaturated systems, though can manifest in other configurations as well.
Static, ground state deformation effects of the tensor force can also affect cross sections by shifting the fusion barrier.

\item[Conclusions]
The tensor interaction has a measurable effect on the fusion cross sections of nuclei spanning the nuclear chart.
The effect comes from both static effects present in the ground state and dynamic processes arising from the time evolution of the system.
This motivates the development of a modern Skyrme parameter set that includes all time-odd and tensor terms and that studies moving forward should include the tensor force to ensure a more robust and complete description of nuclei.

\end{description}
\end{abstract}


\maketitle
\end{CJK*}
\section{Introduction}
\label{introduction}
Many applications of nuclear physics rely on our description of both the structure and dynamics of exotic nuclei.
From superheavy and neutron rich nuclei formation to modeling the rapid neutron capture process (r-process), the fundamental description of static nuclei and how they interact is of great importance to accurately inform emergent theories.
In the current era with state-of-the-art radioactive ion beams becoming more available we are presented with excellent opportunities in both the experimental and theoretical study of exotic nuclei and their interactions~\cite{balantekin2014}.

It is with this motivation that the influence of the tensor interaction on fusion cross sections is being studied.
While the importance of the tensor force has been well studied in nuclear structure calculations~\cite{otsuka2006,lesinski2007,colo2007,bai2010}, the impact on nuclear dynamics has only recently been a topic of interest.
This relative neglect is primarily due to the tensor interaction not being included in most studies of nuclear reactions, even though the tensor force plays a significant role in intrinsic excitations that may introduce unforeseen dynamical effects.
The inclusion of these effects provide a more complete picture into the effective nucleon-nucleon interaction and nuclear reactions in general.

The study of fusion cross sections both above and below the barrier has been performed for many systems using a number of distinct techniques, though most approaches ultimately require a heavy-ion interaction potential as the starting point~\cite{back2014}.
How one obtains such a potential is also varied, though two main classes can be roughly identified: phenomenological models~\cite{bass1974,randrup1978a,satchler1979,rhoadesbrown1983a,seiwert1984,adamian2004,wang2012,zhu2016,bao2016,feng2017} and (semi-)microscopic models~\cite{brueckner1968,moller2004,guo2004,guo2005,misicu2006,umar2006b,misicu2007,guo2007b,lu2014,simenel2017}.
Within each class there are myriad methods and assumptions, so we focus on the (semi-)microscopic class of methods which are more germane to the current work.
One reason for pursuing a microscopic approach to describe fusion is the desire to have the theoretical description be as close to the underlying physical processes as possible in hopes that this will produce a more predictive technique.
It was this desire that led to techniques such as the density-constrained time-dependent Hartree-Fock (DC-TDHF) approach~\cite{umar2006b} for calculating heavy-ion interaction potentials which naturally incorporates all dynamical effects coming from the TDHF description of the collision process~\cite{simenel2018}.
By using a time evolution of the heavy-ion system as the starting point, you avoid the unphysically large density overlaps seen in simple folding model methods and obtain dynamical effects not seen otherwise.
These effects include nucleon transfer, neck formation, internal excitations, and deformation effects and manifest from the time evolution of the initial configuration.
That is to say, the dynamic densities evolve self-consistently in a fully microscopic description of two boosted nuclei with no outside constraints.
The density-constraint that is performed to the evolved densities is what provides the static collective energy which is then interpreted as the ion-ion interaction potential.
The DC-TDHF approach has been applied to many systems spanning the nuclear chart and the fusion cross section results generally agree well with available experimental data~\cite{umar2006a,umar2006d,umar2008b,umar2008a,umar2009b,umar2010c,oberacker2010,keser2012,umar2012a,umar2014a,godbey2017,godbey2019b}.

By themselves, direct TDHF calculations have been used in recent years to investigate the impact of the tensor force on heavy-ion collisions~\cite{fracasso2012,dai2014a,stevenson2016,shi2017,guo2018}, though no studies employing phenomenological nor microscopic methods for obtaining ion-ion potentials have included the tensor interaction until recently~\cite{guo2018b}.
To that end, the current work is a follow-up to the initial investigation of fusion barriers for a wide variety of nuclear systems and the role the tensor force plays in building those potentials.
A more detailed description of the methods and theory are in Sec.~\ref{theory}, fusion cross section results are presented and discussed in Sec.~\ref{results}, and a brief summary and conclusion comprises Sec.~\ref{summary}. 

\section{Theoretical framework}
\label{theory}

While TDHF has been used for a vast set of problems throughout its history, there is not one standard approach that is universally followed.
From the choice of geometric symmetries to the particular EDF employed, there are potential assumptions or omissions that could then lead to an incomplete (or inaccurate) description of nuclear dynamics.
The effective interaction in particular is one area which has been a subject of study since the theory's inception.
One historical example concerns the time-odd terms of the energy density functional (EDF) which are shown to be non-negligible in heavy-ion collisions~\cite{umar2006c}.
This provides a motivating analogy to the tensor force between nucleons, as it has only recently been included in TDHF investigations as discussed above and may be similarly important.

\subsection{Full Skyrme energy functional}

The bulk of TDHF calculations utilize the Skyrme effective interaction~\cite{skyrme1956}, in which the two-body tensor force was proposed in its original form as
\begin{align}
\begin{split}
v_T&=\dfrac{t_\mathrm{e}}{2}\bigg\{\big[3({\sigma}_\mathrm{1}\cdot\mathbf{k}')({\sigma}_\mathrm{2}\cdot\mathbf{k}')-({\sigma}_\mathrm{1}\cdot{\sigma}_\mathrm{2})\mathbf{k}'^{\mathrm{2}}\big]\delta(\mathbf{r}_\mathrm{1}-\mathbf{r}_\mathrm{2})\\
&+\delta(\mathbf{r}_\mathrm{1}-\mathbf{r}_\mathrm{2})\big[3({\sigma}_\mathrm{1}\cdot\mathbf{k})({\sigma}_\mathrm{2}\cdot\mathbf{k})-({\sigma}_\mathrm{1}\cdot{\sigma}_\mathrm{2})\mathbf{k}^\mathrm{2}\big]\bigg\}\\
&+t_\mathrm{o}\bigg\{3({\sigma}_\mathrm{1}\cdot\mathbf{k}')\delta(\mathbf{r}_\mathrm{1}-\mathbf{r}_\mathrm{2})({\sigma}_\mathrm{2}\cdot\mathbf{k})-({\sigma}_\mathrm{1}\cdot{\sigma}_\mathrm{2})\mathbf{k}'
\delta(\mathbf{r}_\mathrm{1}-\mathbf{r}_\mathrm{2})\mathbf{k}\bigg\}.
\end{split}
\end{align}
The coupling constants $t_\textrm{e}$ and $t_\textrm{o}$ represent the strengths of triplet-even and
triplet-odd tensor interactions, respectively.  The operator $\mathbf{k}=\frac{1}{2i}(\nabla_1-\nabla_2)$ acts on the right and
$\mathbf{k}'=-\frac{1}{2i}(\nabla'_1-\nabla'_2)$ acts on the left.

The total energy of the system may then be represented as
\begin{equation}
E=\int d^3r  {\cal H}(\rho, \tau, {\mathbf{j}}, {\textbf{s}}, {\textbf{T}}, {\textbf{F}}, J_{\mu\nu}; {\mathbf{r}})
\label{Energy}
\end{equation}
in terms of the energy functional. The functional is composed by
the number density $\rho$, kinetic density $\tau$, current density ${\mathbf{j}}$, spin density ${\mathbf{s}}$, spin-kinetic density ${\mathbf{T}}$,
the tensor-kinetic density {\textbf{F}}, and spin-current pseudotensor density $J$~\cite{stevenson2016}. The full version of Skyrme EDF is expressed as
\begin{align}
\label{EDFH}
\begin{split}
\mathcal{H}&=\mathcal{H}_0+\sum_{\rm{t=0,1}}\Big\{A_{\rm{t}}^{\rm{s}}\mathbf{s}_{\rm{t}}^2+(A_{\rm{t}}^{\Delta{s}}+B_{\rm{t}}^{\Delta{s}})
\mathbf{s}_{\rm{t}}\cdot\Delta\mathbf{s}_{\rm{t}}+B_{\rm{t}}^{\nabla s}(\nabla\cdot \mathbf{s}_{\rm{t}})^2 \\
&+B_{\rm{t}}^{F}\big(\mathbf{s}_{\rm{t}}\cdot
\mathbf {F}_{\rm{t}}-\frac{1}{2}\big(\sum_{\mu=x}^{z}J_{\rm{t}, \mu\mu}\big)^2-\frac{1}{2}\sum_{\mu, \nu=x}^{z}J_{\rm{t}, \mu\nu}J_{\rm{t}, \nu \mu}\big)\\
&+(A_{\rm{t}}^{\rm{T}}+B_{\rm{t}}^{\rm{T}})\big(\mathbf{s}_{\rm{t}}\cdot\mathbf{T}_{\rm{t}}-
\sum_{\mu,\nu=x}^{z}J_{\rm{t},\mu\nu}J_{\rm{t},\mu\nu}\big)\Big\},
\end{split}
\end{align}
where $\mathcal{H}_0$ is the simplified functional used in the standard Sky3D TDHF code~\cite{maruhn2014}.
The terms containing the coupling constants $A$ arise from the Skyrme central force and those with $B$ from the tensor force. The definitions of the constants $A$ and $B$
can be found in Refs.~\cite{lesinski2007,davesne2009}. All of the time-even and time-odd terms in Eq.~(\ref{EDFH}) have been implemented numerically in the mean-field Hamiltonians of the HF, TDHF, and DC-TDHF approaches.
As discussed in Refs.~\cite{lesinski2007,stevenson2016}, the terms containing the gradient of the spin density (the $\mathbf{s}_{\rm{t}}\cdot\Delta\mathbf{s}_{\rm{t}}$ and $(\nabla\cdot \mathbf{s}_{\rm{t}})^2$ terms) may cause spin instability in both nuclear structure and reaction studies, and are thus turned off in these calculations.
This omission of the spin density gradient terms is standard procedure in TDHF studies that include the time-odd terms.

\subsection{TDHF approach}

Given a many-body Hamiltonian, the action can be constructed as
\begin{equation}
S=\int_{t_1}^{t_2} dt \langle \Phi(\textbf{r}, t)|H-i\hbar \partial_t|\Phi(\textbf{r}, t)\rangle ,
\end{equation}
where $\Phi$ is the time-dependent many-body wave function.
In TDHF approach the many-body wave function $\Phi(\mathbf{r}, t)$ is approximated as a single time-dependent Slater determinant composed of an antisymmetrized product
of the single particle states $\phi_{\rm{\lambda}}(\mathbf{r}, t)$
\begin{equation}
\Phi(\mathbf{r}, t)=\frac{1}{\sqrt{N!}}\textrm{det}\{\phi_{\rm{\lambda}}(\mathbf{r}, t)\},
\end{equation}
and this form is kept at all times in the dynamical evolution.
By taking the variation of the action with respect to the single-particle wave functions, the set of nonlinear coupled TDHF equations in the multidimensional
space-time phase space
\begin{equation}
i\hbar \frac{\partial}{\partial_t}\phi_{\rm{\lambda}}(\mathbf{r}, t)=h\phi_{\rm{\lambda}}(\mathbf{r}, t)
\end{equation}
yields the most probable time-dependent mean-field path, where $h$ is the HF single-particle Hamiltonian.
For this study, the three-dimensional coordinate space TDHF code Sky3D~\cite{maruhn2014} has been extended to include the tensor terms of the interaction.
To expand on the numerical details of the simulations, the static HF ground state for the reaction partner has been calculated on the symmetry-unrestricted three-dimensional grid.
The resulting Slater determinants for each nucleus are combined and comprise the larger Slater determinant describing the colliding system.
The TDHF time propagation is performed using a Taylor-series expansion up to the sixth order of the unitary boost operator with a time step of $0.2 ~\mathrm {fm}/c$.
For the dynamical evolution, we use a numerical box of 48 fm along the collision axis and 24 fm in the other two directions and a grid spacing of 1.0~fm.
The initial separation between the two nuclei is 20~fm.

\subsection{Dynamical potential from DC-TDHF approach}

The semi-classical nature of TDHF precludes one from fully describing heavy-ion fusion from TDHF calculations alone; while direct TDHF can provide fusion cross sections that agree well above the barrier, sub-barrier calculations result in scattering due to the absence of a many-body description of quantum tunneling.
At present, all sub-barrier fusion calculations assume that there exists an ion-ion potential which depends on the internuclear distance.
The microscopic DC-TDHF approach~\cite{umar2006b} is employed to extract this heavy-ion potential from the TDHF time evolution of the dinuclear system at above-barrier energies.
This approach allows for the determination of an interaction potential that includes all dynamical effects seen in the TDHF evolution.
To construct this potential, at certain times during the evolution, the instantaneous TDHF density is used to perform a static HF energy minimization
\begin{equation}
\delta \langle \Psi_{\rm DC}|H-\int d^3r \lambda(\textbf{r})\rho(\textbf{r})|\Psi_{\rm DC}\rangle=0,
\end{equation}
by constraining the proton and neutron densities to be equal to the instantaneous TDHF densities. As it is the total density that is being constrained, all moments are simultaneously constrained.
DC-TDHF calculations give the adiabatic reference state for a given TDHF state, which is the Slater determinant with the lowest energy for a given density.
The minimized energy
\begin{equation}
E_{\rm {DC}}(\textbf R)=\langle \Psi_{\rm DC}|H|\Psi_{\rm DC}\rangle
\end{equation}
is the so-called density-constrained energy.
Since this density-constrained energy still contains the binding energies of the initial individual nuclei, the heavy-ion interaction potential is deduced as
\begin{equation}
V(\textbf R)=E_{\rm {DC}}(\textbf R)-E_{\rm {1}}-E_{\rm {2}},
\label{VB}
\end{equation}
where $E_{\rm {1}}$ and $E_{\rm {2}}$ are the binding energies of the two individual nuclei.
This procedure is performed separately during the time evolution and does not affect the TDHF time evolution in any way.
The result is a microscopically determined ion-ion interaction potential that contains no free parameters or normalization.
We have performed the density constraint calculations at every 10-20 time steps (corresponding to 2-4~$\mathrm {fm}/c$ interval).
The choice on how often to perform the density constraint depends entirely on the resolution you want to obtain for the potential and does not affect the evolution in any way.
The convergence criteria in DC-TDHF calculations is as good if not better than in the traditional constrained HF with a constraint on a single collective degree of freedom.

\subsection{Calculation of cross sections from interaction potential}\label{sec:cs}

As described above, DC-TDHF provides a way to obtain one dimensional ion-ion fusion potentials, which can then be used to calculate fusion cross sections.
The procedure to obtain transmission probabilities $T_{l}(E_{\mathrm{c.m.}})$ (and thus cross sections) from an arbitrary one-dimensional potential can solved by numerical integration of the two-body Schr\"odinger equation:

\begin{equation}
\left[\frac{-\hbar^2}{2\mathrm{M(R)}}\frac{d^2}{dR^2}+\frac{l(l+1)\hbar^2}{2\mathrm{M(R)}R^2} + V(R) - E\right]\psi=0 .
\label{eq:se}
\end{equation}

The incoming wave boundary conditions (IWBC) method is used to calculate transmission probabilities which assumes that fusion occurs once the minimum of V(R) is reached~\cite{rawitscher1964}.
The subtle assumption here is that all transmitted waves (or waves with an energy greater than the barrier) will lead to fusion.
This assumption is the primary issue that complicates descriptions of light nuclei systems like $^{12}$C+$^{12}$C as discussed in~\cite{jiang2013,godbey2019b}.

The barrier penetrability $T_l(E_{\mathrm{c.m.}})$ is then the ratio of the incoming flux at the minimum of the potential inside the barrier to the incoming flux at a large distance.
Once $T_l(E_{\mathrm{c.m.}})$ is calculated, the fusion cross sections at energies above and below the barrier are calculated as
\begin{equation}
\sigma_f(E_{\mathrm{c.m.}})=\frac{\pi}{k_0^2}\sum_{l=0}^{\infty}(2l+1)T_l(E_{\mathrm{c.m.}}).
\end{equation}

As DC-TDHF potentials are the result of a TDHF evolution, the coordinate-dependent mass $M(R)$ can
be calculated directly from TDHF dynamics~\cite{umar2009a,umar2009b}.
This mass primarily influences the inner part of the barrier, leading to a broader
barrier width thus leading to further hindrance in the sub-barrier region.
The effect of the coordinate-dependent mass also plays a role in the energy dependence of the potential~\cite{umar2014a}, as the value of the mass will spike as the nuclei slow down at the point of the barrier.
Instead of solving the Schr\"odinger equation using the coordinate-dependent mass M(R), the potential can be transformed by a scale factor~\cite{umar2009b,goeke1983}
\begin{equation}
d\bar{R}=\left(\frac{\mathrm{M(R)}}{\mu}\right)^{\frac{1}{2}}dR.
\end{equation}
Upon making this transformation the coordinate-dependence of M(R) vanishes and is replaced by the reduced mass $\mu$ in Eq.~(\ref{eq:se}) and the Schr\"odinger equation is solved using the modified Numerov method as it is formulated in the coupled-channel code CCFULL~\cite{hagino1999}.

\section{Results}
\label{results}


In choosing representative systems to investigate the effect of the tensor interaction on fusion cross sections, a variety of nuclei were chosen from multiple mass regions and spin structures in an attempt to characterize what effects could be expected from the tensor force's inclusion.
One point of interest is in the asymmetry of the interacting fragments, as phenomena like neck formation will take on a more pronounced role for heavily mass asymmetric systems.
Additionally, isovector contributions to the potential will be lessened in symmetric collisions (such as $^{48}\mathrm{Ca}+\mathrm{^{48}Ca}$), and the tensor contribution will be primarily from the isoscalar part of the EDF~\cite{godbey2017}.

A note should be made on the Skyrme functionals used in this work, in which the tensor force has been constructed in two ways.
One is to add the force perturbatively to an existing standard interaction as was done with the Skyrme parameter set SLy5~\cite{chabanat1998a}, where the tensor force was added and refit to data, resulting in SLy5t~\cite{colo2007}.
The comparison between calculations with SLy5 and SLy5t addresses the question on how much of the changes is caused by tensor force itself, as only the tensor parameters were refit.
Another approach is to readjust the full set of Skyrme parameters self-consistently.
This strategy has been adopted in Ref.~\cite{lesinski2007} and led to the set of T$IJ$ parametrizations with a wide range of isoscalar and isovector tensor couplings.
Due to its fitting strategy, the contributions from the tensor force and all other terms cannot so easily be separated.

\begin{figure}
	\includegraphics[width=8.6cm]{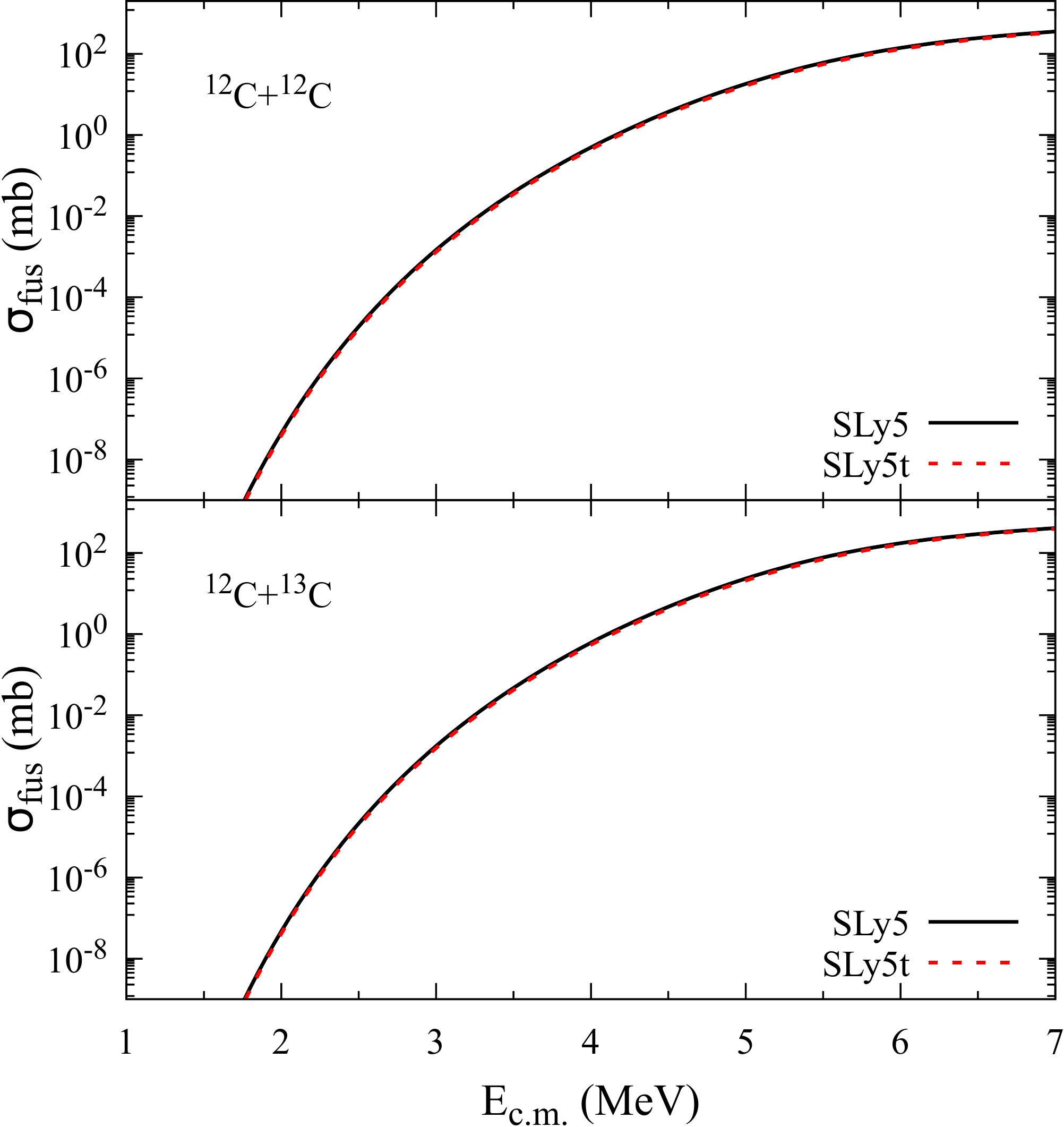}
	\caption{(Color online) Fusion cross sections obtained from the DC-TDHF approach for $^{12}\mathrm{C}+\mathrm{^{12}C}$ at $E_{\mathrm{c.m.}}=8$~MeV and $^{12}\mathrm{C}+\mathrm{^{13}C}$ at $E_{\mathrm{c.m.}}=8$~MeV with SLy5 (black solid line) and SLy5t (red dashed line) forces.
		\label{Fig:CCxsec}}
\end{figure}

Starting with light systems, fusion cross section results from SLy5 and SLy5t are shown in Fig.~\ref{Fig:CCxsec} for the spin-unsaturated $^{12}\mathrm{C}+\mathrm{^{12}C}$ and $^{12}\mathrm{C}+\mathrm{^{13}C}$.
For both SLy5 and SLy5t spherical ground state solutions were found for both $^{12}\mathrm{C}$ and $^{13}\mathrm{C}$.
In Ref.~\cite{guo2018b}, a small difference in interaction potentials was found for $^{12}\mathrm{C}+\mathrm{^{12}C}$, though that difference does not result in an appreciable variation of the cross sections at any energy above or below the barrier at the scales used here.

\begin{figure}
	\includegraphics[width=8.6cm]{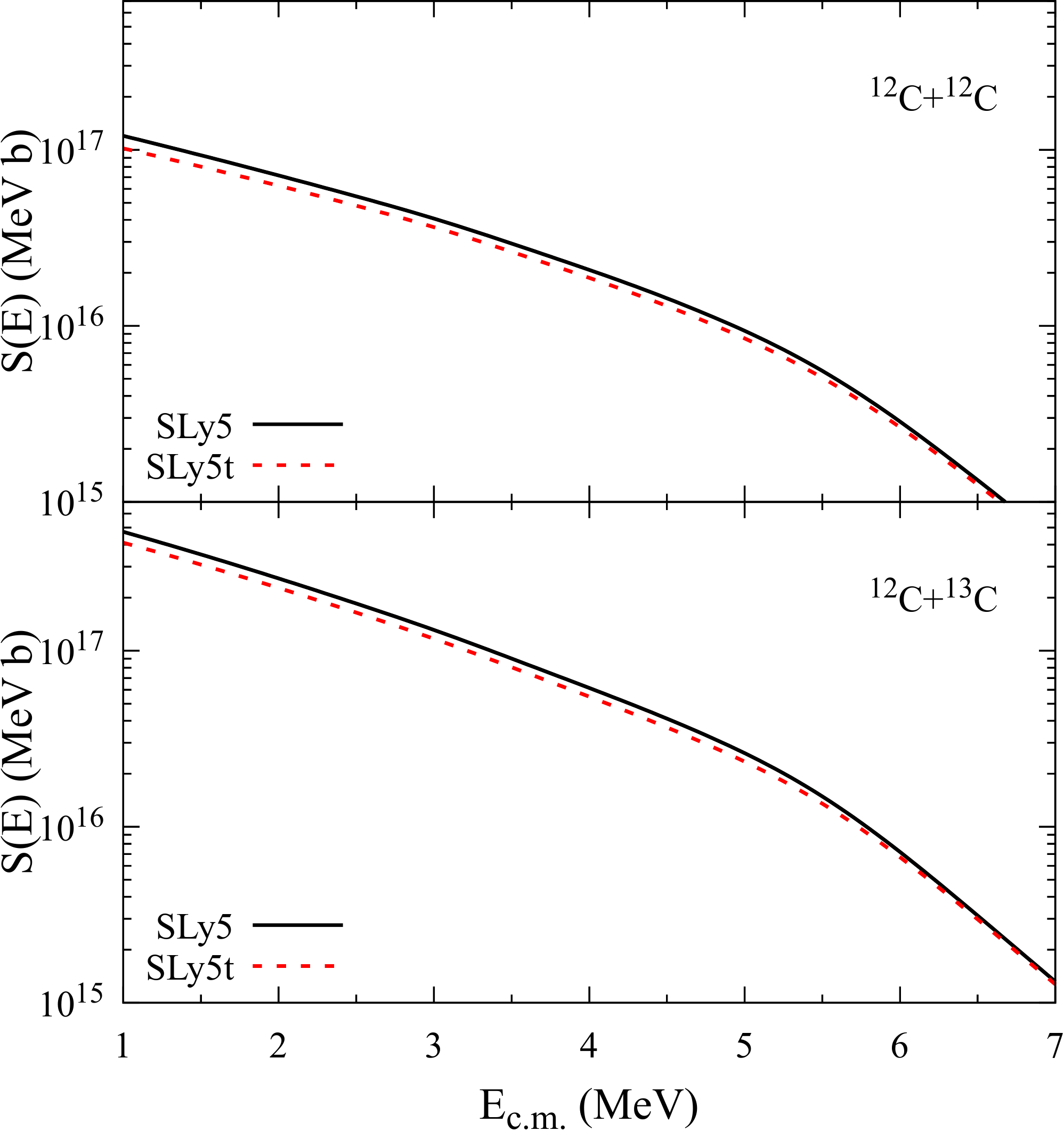}
	\caption{(Color online) $S$ factors obtained from the DC-TDHF approach for $^{12}\mathrm{C}+\mathrm{^{12}C}$ at $E_{\mathrm{c.m.}}=8$~MeV and $^{12}\mathrm{C}+\mathrm{^{13}C}$ at $E_{\mathrm{c.m.}}=8$~MeV with SLy5 (black solid line) and SLy5t (red dashed line) forces.
		\label{Fig:CCsfac}}
\end{figure}

$^{12}\mathrm{C}+\mathrm{^{12}C}$ plays a significant role in nucleosynthesis and the r-process and thus the energies of focus are often in the extreme sub-barrier regime ($\approx 1$~MeV)~\cite{cumming2001,strohmayer2002,hoyle1954,godbey2019b}.
Figure~\ref{Fig:CCxsec} spans the entire sub-barrier region, though the large range of values prohibits a close comparison of extreme sub-barrier values.
This is a deficiency of logarithmic cross section plots which is addressed in most studies of light systems of astrophysical interest by instead plotting the $S$~factor
\begin{equation}
S(E_{\mathrm{c.m.}})=\sigma(E_{\mathrm{c.m.}})E_{\mathrm{c.m.}}e^{2\pi\eta},
\end{equation}
Where $E_{\mathrm{c.m.}}$ is the center of mass energy, $\eta=\mathrm{Z_1}\mathrm{Z_2}e^2/\hbar\nu$ is the Sommerfeld parameter, and $\nu$ is the relative velocity of the nuclei at infinity.
The $S$~factor is used primarily due to it removing most of the energy dependence at very low energies, permitting a closer comparison of data and experiment.
This quantity is presented in Fig.~\ref{Fig:CCsfac} for both $^{12}\mathrm{C}+\mathrm{^{12}C}$ and $^{12}\mathrm{C}+\mathrm{^{13}C}$.
In contrast to the cross section results, a small difference can be discerned at sub-barrier energies which slightly increases as one descends.
The small decrease seen here is of note, though the tensor interaction does not change the overall structure of the $S$ factor at sub-barrier energies via either the manifestation of resonant structures seen in the experimental data or in strong hindrance effects which have been suggested to be present in this system~\cite{jiang2007}.

\begin{figure}
	\includegraphics[width=8.6cm]{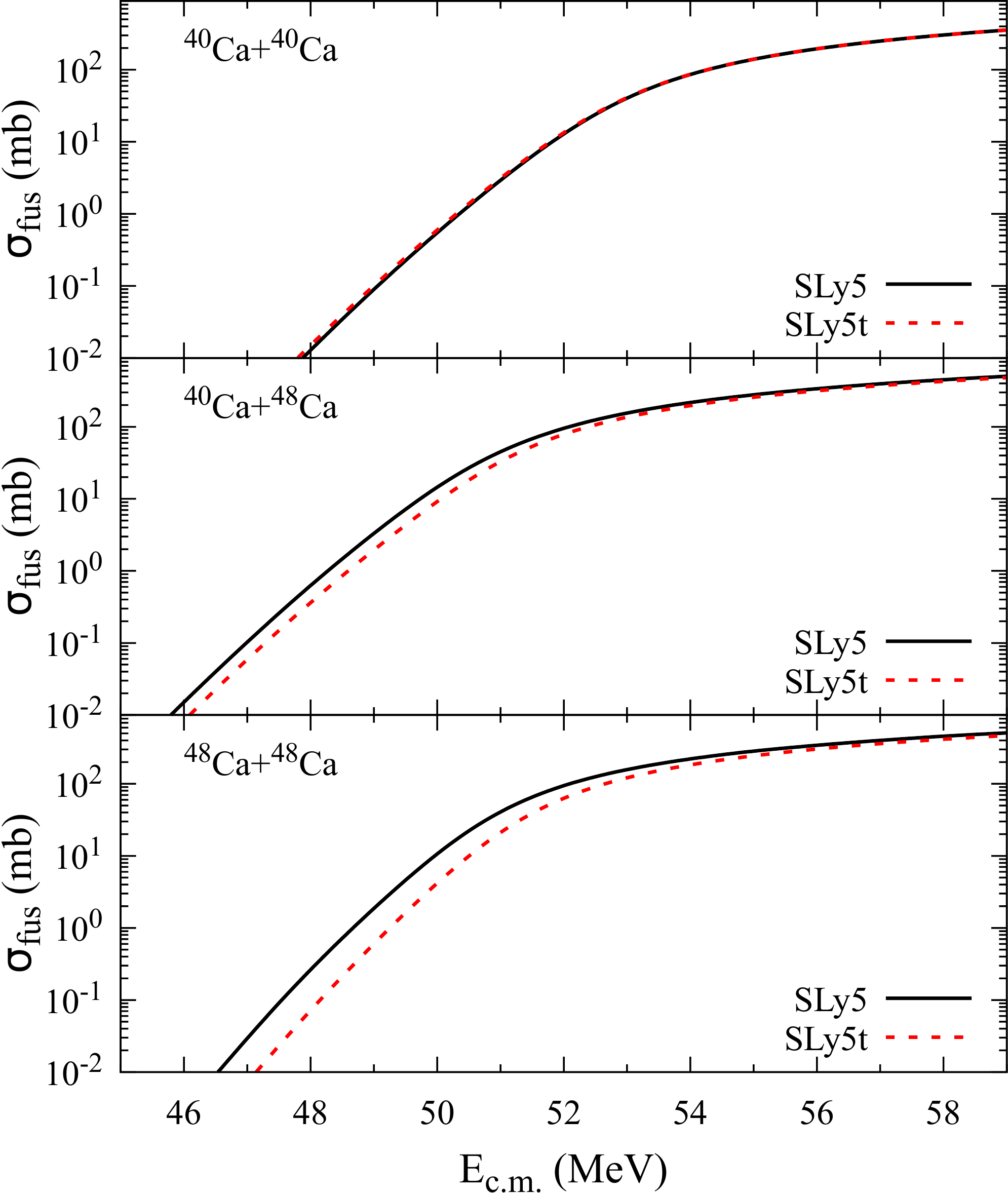}
	\caption{(Color online) Fusion cross sections obtained from the DC-TDHF approach for $^{40}\mathrm{Ca}+\mathrm{^{40}Ca}$ at $E_{\mathrm{c.m.}}=55$~MeV, $^{40}\mathrm{Ca}+\mathrm{^{48}Ca}$ at $E_{\mathrm{c.m.}}=55$~MeV, and
	$^{48}\mathrm{Ca}+\mathrm{^{48}Ca}$ at $E_{\mathrm{c.m.}}=55$~MeV with SLy5 (black solid line) and SLy5t (red dashed line) forces.
		\label{Fig:CaCaxsec}}
\end{figure}

Moving up in mass, Fig.~\ref{Fig:CaCaxsec} presents cross sections obtained for $^{40}\mathrm{Ca}+\mathrm{^{40}Ca}$, $^{40}\mathrm{Ca}+\mathrm{^{48}Ca}$, and $^{48}\mathrm{Ca}+\mathrm{^{48}Ca}$.
These systems are valuable benchmarks as both $^{40}\mathrm{Ca}$ and $\mathrm{^{48}Ca}$ are closed-shell nuclei, though $\mathrm{^{48}Ca}$ has $28$ neutrons, corresponding to the spin-unsaturated magic number 28.
The symmetric spin-saturated $^{40}\mathrm{Ca}+\mathrm{^{40}Ca}$ shows almost no difference when comparing the fusion cross sections obtained using the SLy5 and SLy5t forces, though there is a slight variation at low energies.
The finding that SLy5 and SL5t result in nearly identical cross sections at all energy regimes suggests that, even during a dynamic evolution at energies above the barrier, the tensor force does not play a substantial role in collisions between spin-saturated systems.
Moving to the asymmetric $^{40}\mathrm{Ca}+\mathrm{^{48}Ca}$ reaction presents the first example of how the tensor force results in different behavior in the sub-barrier region, with the tensor interaction resulting in a slight fusion hindrance.
The difference is small in this case, and it is only substantially different at energies below the barrier.
Finally, fusion cross sections of $^{48}\mathrm{Ca}+\mathrm{^{48}Ca}$ are investigated with and without the inclusion of the tensor interaction.
It is in this symmetric system that the largest effect can be seen for the series of calcium reactions.
As mentioned above, both fragments have spin-unsaturated neutron shells which is likely the reason for the greater contribution of the tensor force to further fusion hindrance in the sub-barrier region.

\begin{figure}
	\includegraphics[width=8.6cm]{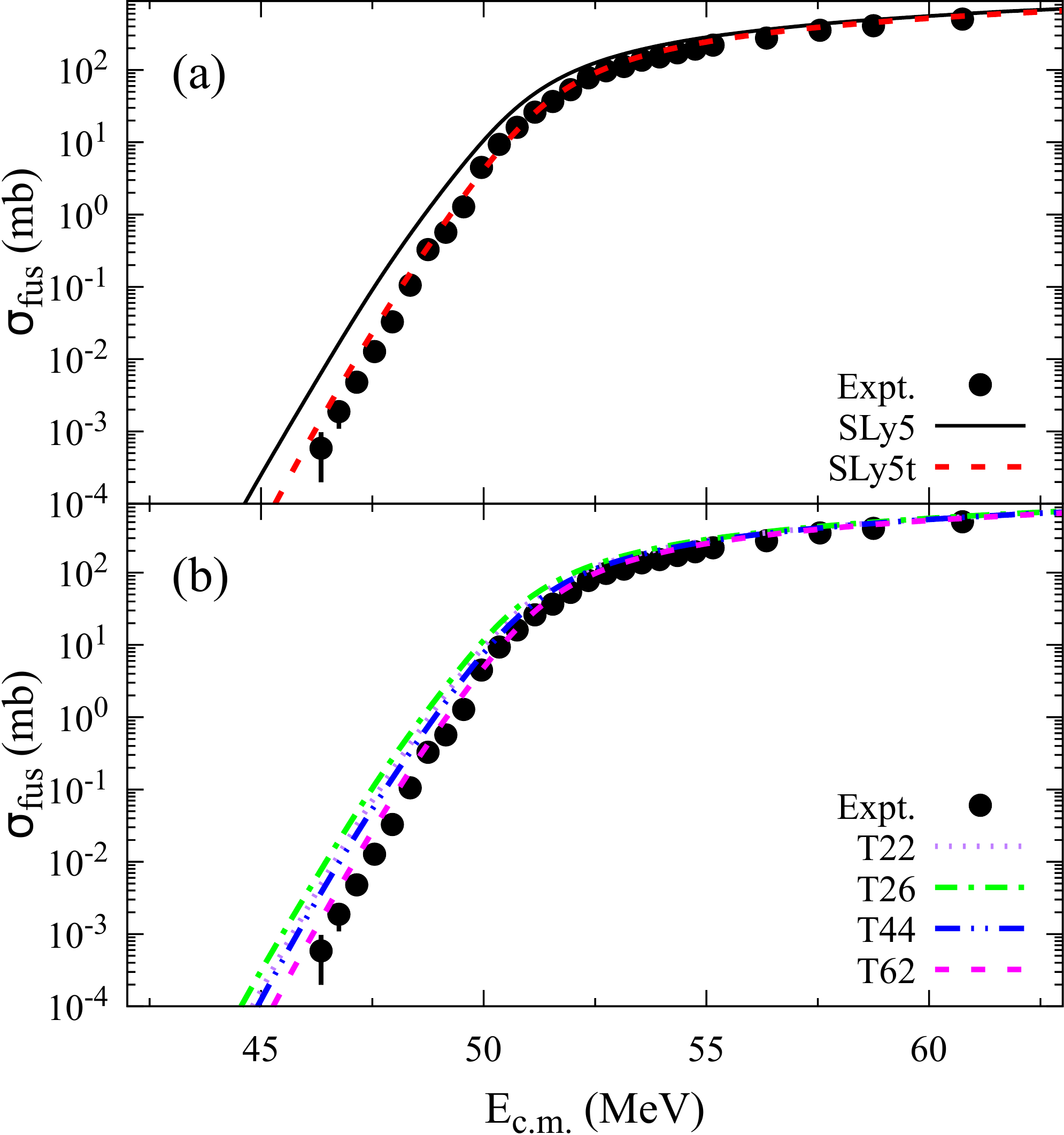}
	\caption{(Color online) Fusion cross sections obtained from the DC-TDHF approach and experimental data from~\cite{stefanini2009} for $^{48}\mathrm{Ca}+\mathrm{^{48}Ca}$ at $E_{\mathrm{c.m.}}=55$~MeV with (a) SLy5 (black solid line) and SLy5t (red dashed line) forces and (b) T22 (purple dotted line), T26 (green dash-dotted line), T44 (blue dash-dot-dotted line), and T62 (orange dashed line) forces.
	\label{Fig:CaTXXxsec}}
\end{figure}

As the $^{48}\mathrm{Ca}+\mathrm{^{48}Ca}$ results exhibit a large effect when the tensor force is added, it is an excellent candidate system to compare with experimental data for fusion cross sections.
Such a comparison is shown in Fig.~\ref{Fig:CaTXXxsec} for both SLy5(t) and a selection of the T$IJ$ forces.
Starting first in panel (a), it is clear that the inclusion of the tensor interaction corrects the overestimation present when using SLy5 primarily in the sub-barrier region, though the area around the barrier itself is also more in line with what is seen experimentally.
To investigate the source of this correction, it is useful to compare different methods of including the tensor interaction as a check for consistency.
To this end a selection from the T$IJ$ set of forces was also used; particularly the T22, T26, T44, and T62 forces which each have different isoscalar and isovector couplings.
Panel (b) of Fig.~\ref{Fig:CaTXXxsec} plots these results from the T$IJ$ set of forces and a similar correction is seen for the T62 interaction.
The other predicted cross sections seem to have quite a varied behavior in the sub-barrier region, suggesting that isoscalar and isovector couplings play a role in driving the dynamics for this reaction.
The two forces that best reproduced the data were SLy5t and T62 which have very different values for the spin-current coupling constants, though this could perhaps be explained by the fact that SLy5t had these constants determined as an addition to the unchanged SLy5 functional.
Regardless, it is clear that for medium mass, spin-unsaturated systems much is gained by including the tensor interaction in the calculations.

\begin{figure}
	\includegraphics[width=8.6cm]{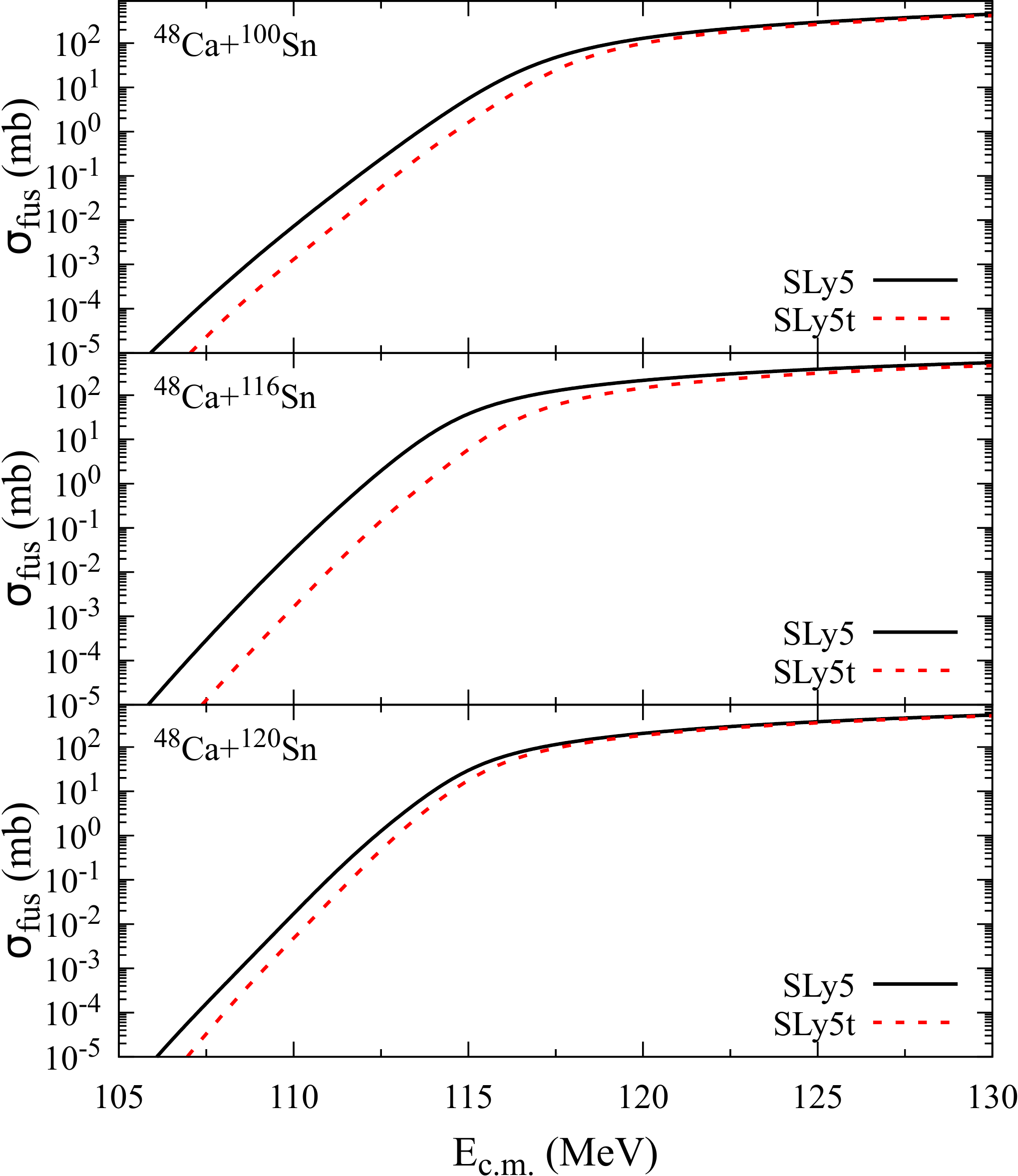}
	\caption{(Color online) Fusion cross sections obtained from the DC-TDHF approach for $^{48}\mathrm{Ca}+\mathrm{^{100}Sn}$ at $E_{\mathrm{c.m.}}=125$~MeV, $^{48}\mathrm{Ca}+\mathrm{^{116}Sn}$ at $E_{\mathrm{c.m.}}=125$~MeV, and	$^{48}\mathrm{Ca}+\mathrm{^{120}Sn}$ at $E_{\mathrm{c.m.}}=125$~MeV with SLy5 (black solid line) and SLy5t (red dashed line) forces.
		\label{Fig:CaSnxsec}}
\end{figure}

Next we consider the $^{48}\mathrm{Ca}+\mathrm{^{100,116,120}Sn}$ series which represents a set of mixed-mass systems comprised of a heavy-mass target and a medium-mass projectile.
By increasing the mass asymmetry one expects to see larger rearrangement during the reaction resulting in more particle transfer and a more dynamic neck formation around the peak of the barrier, all of which may affect the interaction potential.
The results for both SLy5 and SLy5t are presented in Fig.~\ref{Fig:CaSnxsec}.
For all systems, a deviation from the non-tensor results is observed, though the magnitude varies between them.
It should be noted that for $^{48}\mathrm{Ca}$, $^{100}\mathrm{Sn}$, and $^{120}\mathrm{Sn}$ the ground state solutions for both forces are found to be spherical, though $^{116}\mathrm{Sn}$ has a slight quadrupole deformation.
As the reaction dynamics depend strongly on the initial configuration of the fragments, the symmetry axis of $^{116}\mathrm{Sn}$ is aligned parallel to the collision axis for both SLy5 and SLy5t.
The deformation is reduced in SLy5t ($\beta_2=0.026$ as opposed to $\beta_2=0.077$ seen in SLy5), and thus part of the difference seen for this nucleus could arise from static effects originating in the initial ground state configuration.
The reduced deformation is due to the inclusion of the tensor interaction, however, and thus the deviation in cross sections can still be compared.

\begin{figure}
	\includegraphics[width=8.6cm]{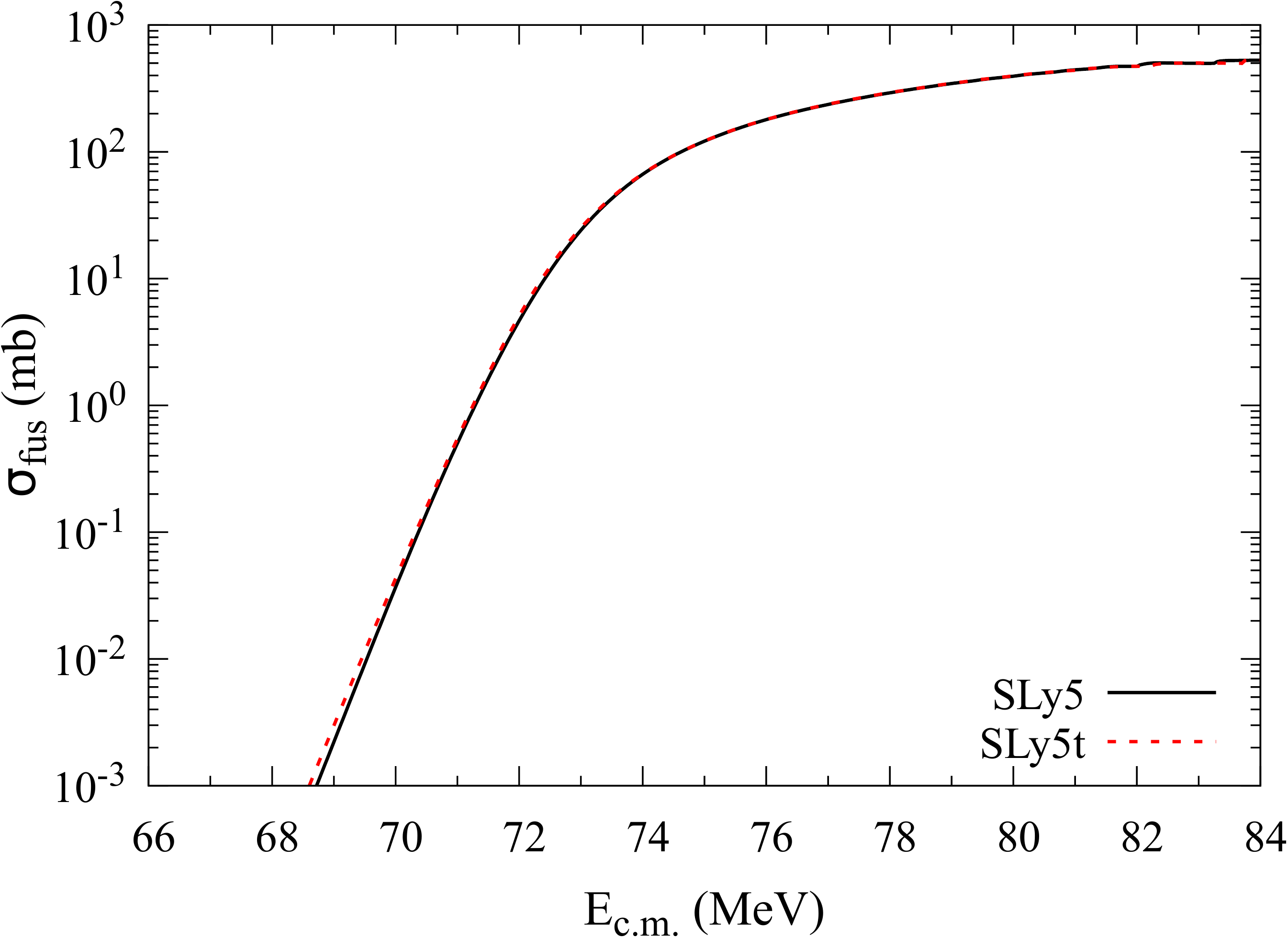}
	\caption{(Color online) Fusion cross sections obtained from the DC-TDHF approach for $^{16}\mathrm{O}+\mathrm{^{208}Pb}$ at $E_{\mathrm{c.m.}}=75$~MeV with SLy5 (black solid line) and SLy5t (red dashed line) forces.
	\label{Fig:OPbxsec}}
\end{figure}

The final system presented is $^{16}\mathrm{O}+\mathrm{^{208}Pb}$ which further increases the mass asymmetry of the reaction partners to investigate its effect on fusion cross sections.
Both $^{16}\mathrm{O}$ and $\mathrm{^{208}Pb}$ are doubly magic, though only $^{16}\mathrm{O}$ is spin-saturated.
As the mass difference between the two fragments is large, substantial density rearrangement will occur over the course of the time evolution, especially at incoming energies close to the fusion barrier.
As can be seen in Fig.~\ref{Fig:OPbxsec}, however, it appears that complex dynamical rearrangement alone does not significantly involve the tensor interaction for this system.
This also suggests that neck formation and early stage nucleon transfer are not strongly influenced by the inclusion of the tensor force, though comparisons with other systems of nuclei beyond those studied here are needed to confirm this.

\section{Conclusions}

To conclude, the full tensor interaction has been included in a comprehensive study of fusion cross sections using the DC-TDHF technique.
The principal effect of including the tensor force in nuclear reactions is to induce a hindrance effect in the sub-barrier energy regime, though the magnitude of the effect is not uniform for all systems.
For example, light systems only see an effect at extreme sub-barrier energies.
Even then, the effect is limited to a slight changing of the slope of the cross sections.
The move to medium and heavy mass systems on the other hand sees a noted deviation in fusion cross sections when comparing interactions with and without the tensor interaction.
Spin-unsaturated systems in particular experience the largest effect and can bring cross sections results more in line with experimental data when the tensor terms are included.

It is clear from these results that the tensor interaction has a measurable effect in nuclear reactions and should thus be included if one desires a more complete description of nuclear processes.
Including the terms in an ad hoc manner as was done in the case of SLy5t appears to give reasonable results, though a full fitting procedure with all terms should be followed to increase confidence in the forces' representative of nuclei.
Parameter sets which have been fit using modern techniques and include all the terms of the Skyrme interaction are well poised to represent the most complete mean-field description of nuclei currently available.

For future studies, further work should of course be done in the investigation of heavy-ion fusion with the full Skyrme interaction.
Heavy systems in particular may be strongly affected by the inclusion of the tensor interaction as well as reactions that lead to super heavy element formation.
In addition to fusion, the influence of the tensor force on quasifission and fission could be studied using the same TDHF codes that have been utilized for this and previous works.

\label{summary}

\section{Acknowledgments}
This work is partly supported by NSF of China (Grants No. 11175252 and 11575189),
NSFC-JSPS International Cooperation Program (Grant No. 11711540016), and Presidential Fund of UCAS,
and by the U.S. Department of Energy under grant No. DE-SC0013847.
The computations in present work have been performed on the High-performance Computing Clusters of SKLTP/ITP-CAS and
Tianhe-1A supercomputer located in the Chinese National Supercomputer Center in Tianjin.

\bibliography{VU_bibtex_master}
\end{document}